\documentclass[aps,prl,reprint,showpacs,footinbib,amsmath,amssymb,amsfonts,floatfix,superscriptaddress]{revtex4-1}

\usepackage{graphicx}
\usepackage{epsfig}
\usepackage{color}

\begin{document}


\title{The Role of Optical Projection in the Analysis of Membrane Fluctuations}



\author{S. Alex Rautu}
\affiliation{Department of Physics, University of Warwick, Coventry, CV4 7AL, United Kingdom}

\author{Davide Orsi}
\affiliation{Department of Physics and Earth Sciences, University of Parma, Italy}

\author{Lorenzo Di Michele}
\affiliation{Cavendish Laboratory, University of Cambridge, Cambridge, CB3 0HE, United Kingdom}

\author{George Rowlands}
\affiliation{Department of Physics, University of Warwick, Coventry, CV4 7AL, United Kingdom}

\author{Pietro Cicuta}
\email{pc245@cam.ac.uk}
\affiliation{Cavendish Laboratory, University of Cambridge, Cambridge, CB3 0HE, United Kingdom}

\author{M. S. Turner}
\email{m.s.turner@warwick.ac.uk}
\affiliation{Department of Physics, University of Warwick, Coventry, CV4 7AL, United Kingdom}


\date{\today}

\begin{abstract}

We propose a methodology to measure the mechanical properties of membranes from their fluctuations and apply this to optical microscopy measurements of giant unilamellar vesicles of lipids. We analyze the effect of the projection of thermal shape undulations across the focal depth of the microscope. We derive an analytical expression for the mode spectrum that varies with the focal depth and accounts for the projection of fluctuations onto the equatorial plane. A comparison of our model with existing approaches, that use only the apparent equatorial fluctuations without averaging out of this plane, reveals a significant and systematic reduction in the inferred value of the bending rigidity. Our results are in full agreement with the values measured through X-ray scattering and other micromechanical manipulation techniques, resolving a long standing discrepancy with these other experimental methods.
\end{abstract}



\maketitle


The optical spectroscopy of thermally induced shape fluctuations (a.k.a.\ flickering) of giant unilamellar vesicles (GUVs) has been used as a method to extract mechanical information about fluid membranes~\cite{Seifert1997, Bassereau2014, Shimobayashi2015}, with similar techniques also applied to red blood cells~\cite{Yoon2009} and other living cells~\cite{Peukes2014}. The most common implementation of this method consists in imaging the equatorial fluctuations via optical microscopy; the fluctuation spectrum, reconstructed by an image analysis, is then compared to a theoretical model that depends on the membrane tension and bending rigidity \cite{Brochard1975, Mutz1990, Meleard1992, Hackl1998, Dobereiner2003, Meleard1998, Faucon1989, Pecreaux2004, Meleard2011, Helfer2000, Henriksen2002, Brown2011}. This has widely been used to assess the differences in the membrane rigidity of various lipid compositions~\cite{Pecreaux2004, Meleard2011, Helfer2000, Henriksen2002, Brown2011}. This quantity controls the physics of membranes, their dynamics~\cite{Seifert1997} and many aspects of lipid mesophase structures~\cite{Komura2014}. Furthermore, it plays a crucial biophysical role in living cells, e.g.\ in cell morphology, motility, and endocytosis~\cite{Lipowsky1995}. 

In this Letter, we demonstrate that current analysis of flickering \cite{Pecreaux2004, Meleard2011} is inadequate. This stems from neglecting the finite focal depth of the microscope, which results in a projection of the shape fluctuations onto the focal region.  Thus, when imaging the equatorial plane of a GUV, the apparent position of the membrane does not correspond to its equatorial contour. Herein, we develop a new methodology to account for this projection. To confirm this procedure and highlight its importance, we perform experiments in which the focal depth can be finely controlled; namely, the GUVs are imaged by a confocal  microscope~\cite{Mertz2009}. We expect our theoretical methodology to be applicable to other imaging techniques, once their focal depths are calibrated. Our analysis is tested on a model system of GUVs made from 1,2-dioleoyl-sn-glycero-3-phosphocholine (DOPC) lipids. This addresses and resolves a longstanding puzzle in the physics of membranes \cite{Nagle2013}: the usual flickering methodology yields an inferred value of the bending rigidity larger ($\sim\!40\%$ for DOPC) than that found by other experimental methods.

\begin{figure}[b!]\includegraphics[width=\columnwidth]{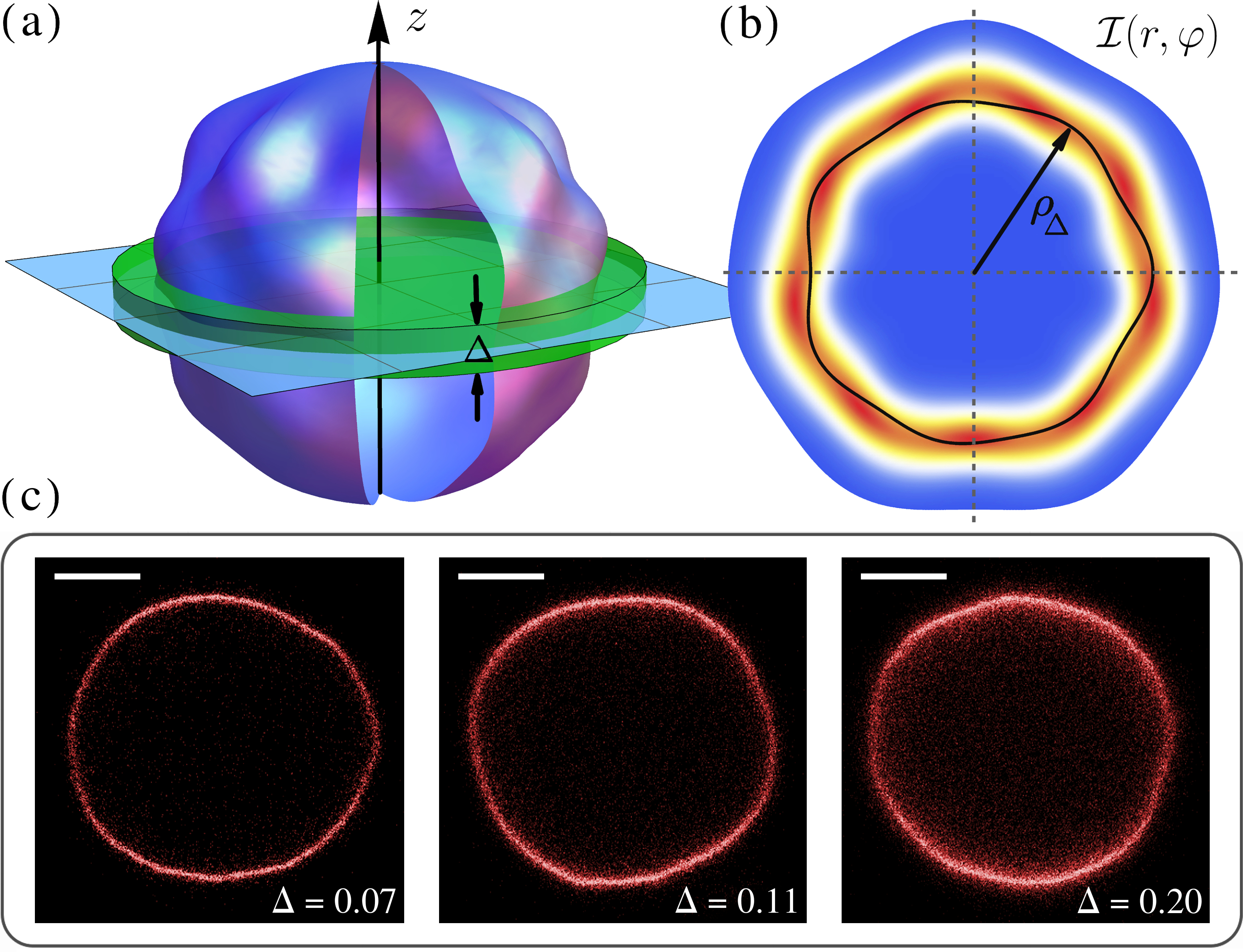}
\caption{\label{fig:1} (color on-line) (a) Schematic diagram of a fluctuating vesicle. Here, the focal plane of the microscope is shown in light blue, while the green slab depicts the region within the focal depth~$\Delta$ (hereinafter, non-dimensionalized by the mean radius of the vesicle), where the surface modes are averaged in projection.
(b) Sketch of a scalar field $\mathcal{I}(r,\varphi)$, corresponding to the membrane density (mass, or intensity for fluorescent membranes) projected onto the focal plane from within the focal depth (red is high density, blue is low). The solid black line represents the first radial moment of $\mathcal{I}(r,\varphi)$, written $\rho_{\scriptscriptstyle\Delta}(\varphi)$.  (c) Confocal fluorescence images of a GUV of mean radius $R=10.2$~$\mu$m at increasing values of $\Delta$. The total intensity increases with $\Delta$ as fluorescence is captured from a greater area of membrane, and the changes in its radial distribution reflect the underlying membrane flickering. The scale bar corresponds to $5$~$\mu$m.
}\end{figure}

The theoretical description of membrane shapes in general, and vesicles in particular, is based on a well established framework~\cite{Canham1970, Helfrich1973, Evans1974}. For undulation wavelengths larger than the membrane thickness, a membrane can be treated as a surface $\mathcal{S}$ with free-energy functional \cite{Seifert1997}
\begin{equation}
  \label{eqn:free-energy}
  \mathcal{F}\left[\mathcal{S}\right]=\int_\mathcal{S}\!\mathrm{d}S\left[\sigma + 2\hspace{1pt}\kappa\left(H-H_0\right)^2 + \bar{\kappa}\hspace{1pt}K\right]\!,
\end{equation} where $\kappa$ is the bending rigidity, $\sigma$ is the surface tension, while $H$ and $H_0$ are the membrane mean curvature and mean spontaneous curvature, respectively. The last term in Eq.~(\ref{eqn:free-energy}), involving the Gaussian curvature $K$ and its associated modulus $\bar{\kappa}$, contributes only as a topologically invariant constant when integrated over a closed surface~$\mathcal{S}$, and it is ignored for the purpose of this study.

Using the quasi-spherical description of vesicles, their membrane surface is given by $\mathcal{S}(\theta,\varphi) =  R \left[1 + u(\theta,\varphi)\right]\hat{\mathbf{r}}$, where $(\theta,\varphi)$  are the spherical angular coordinates, $u(\theta,\varphi)$ is a small deviation about a sphere of radius $R$, and $\hat{\mathbf{r}}$ is the radial unit vector. By using a second-order expansion in $u$, written in the basis of spherical harmonics $Y^m_n$ as $u(\theta,\varphi) = \sum_{n=0}\sum_{\,\left|m\right|\,\leq\,n\,}\mathcal{U}^{m}_{n}\,Y^{m}_{n}(\theta,\varphi)$, where $\mathcal{U}^m_n$ is the mode amplitude, then the mean square deviation of the shape fluctuations is found from Eq.~(\ref{eqn:free-energy}) to be~\cite{Milner1987}
\begin{equation}
  \label{eqn:corr-fct}\big\langle\,\big|\,\mathcal{U}^{m}_{n}\big|^2\big\rangle = \frac{k_B T}{\kappa\,(n-1)(n+2)\left[n(n+1)+\bar{\sigma}\right]}\,,
\end{equation} where the reduced tension $\bar{\sigma} = \sigma\, R^2/\kappa - 2\,H_0 R + 2\,H^2_0 R^2\!$, the mode numbers $n\geq2$ and $\left|m\right|\,\leq\,n$, $k_B$ is the Boltzmann constant, and $T$ is the absolute temperature. 

The standard method of flickering analysis uses Eq.~(\ref{eqn:corr-fct}) to estimate $\kappa$ and $\bar{\sigma}$ from phase contrast or fluorescence imaging of GUVs~\cite{Faucon1989, Pecreaux2004, Meleard2011, Brown2011}. However, these imaging techniques provide only partial information: they give a two-dimensional projection of a three-dimensional fluctuating surface. Typically, the experimental data has been compared to the statistics of the intersection of the vesicle with the focal plane of the objective \cite{Pecreaux2004} at a radial position that we denote by $\rho_0(\varphi, t)$. By projecting the modes in Eq.~(\ref{eqn:corr-fct}) onto the plane $\theta=\pi/2$, and also keeping into account of the average induced in time due to the finite exposure on camera sensors \cite{Pecreaux2004}, the mean square amplitude of the equatorial fluctuations is found to be
\begin{equation}
  \label{eqn:stduq}\left\langle \bar{u}_q(t)\hspace{1pt}\bar{u}^{*}_q(t)\right\rangle_t = \sum_{n\,\geq\,q}\mathcal{E}^2_{n,q}\,\big\langle\!\hspace{1pt}\left|\hspace{1pt}\mathcal{U}^{q}_{n}\right|^2\!\big\rangle\,
 \frac{\tau^{2}_{n}}{\tau^{2}}\!\left(1-e^{-\tau/\tau_{n}}\right)^{\!2}\!\!,
\end{equation}
with $\mathcal{E}_{n,q} = Y^{q}_{n}\!\left(\pi/2,0\right)$ and $\left\langle\cdot\right\rangle_t$ as a time-average~\footnote{See Supplemental Material, which includes Refs. \cite{Seifert1997,Pecreaux2004, Faucon1989, Milner1987, Abramowitz1965, Kardar2007, Henriksen2002, Angelova1999, Sivia2006}, for supporting theoretical calculations, experimental methods, code and example files.}. The coefficients associated with the deviation of the contour from the mean of $\rho_0(\varphi, t)$, a circle of radius $R$, are non-dimensionalized by $R$ and written ${u}_q(t)$. These are experimentally averaged as $\bar{u}_q(t) \equiv \tau^{-1}\!\int^{\tau}_{0}\mathrm{d}t'\,u_q(t+t')$ with $\tau$ the finite acquisition time over which the modes are globally integrated. Also, a viscoelastic theory of the relaxation of a spherical vesicle \cite{Milner1987} is used in Eq.~(\ref{eqn:stduq}) so that $\mathcal{U}_{n,m}(t) = \mathcal{U}_{n,m}(0)\,e^{-t/\tau_{n}}$ where the decay time is
\begin{equation}
  \label{eqn:modelife}
  \tau_n = \frac{ R^3\!\left[\eta_\textnormal{in}\,\frac{(n+2) (2
   n-1)}{n+1}+\eta_\textnormal{out}\,\frac{(n-1) (2 n+3)}{n}\right]}{\kappa \left(n-1\right)\!\left(n+2\right)\left[n (n+1)+\bar{\sigma}\right]},
\end{equation} with $\eta_\textnormal{in}$ and $\eta_\textnormal{out}$ as the viscosities of the surrounding fluid found inside and  outside of the vesicle, respectively.

Although this approach is commonly assumed to be a reasonable approximation, the equatorial cross-section of GUVs is not what is actually observed in microscopy experiments. Strictly speaking, the equator of GUVs provides a vanishing contribution to the image formation both in phase contrast and in fluorescence imaging.  Here, we consider the simpler case of fluorescence emission, and we assume that what is actually observed is a projection over the strip of membrane material that lies within a focal region near the equator, as shown in Fig.~\ref{fig:1}(a). This strip can support a spectrum of surface modes that are averaged out in projection. The effect of this is expected to be particularly strong for modes $q\gtrsim1/\Delta$, where $\Delta$ is the focal depth of the microscope per vesicle radius $R$.

We idealize the acquired optical signal as a convolution of the membrane emission with a Gaussian of width equal to the focal depth. Namely, light arriving from height $z$ above (or below) the focal plane has intensity scaled by $\mathcal{G}(z) = \exp\left\{-z^2/\left[2\,(\Delta R)^2\right]\right\}$. By setting the focal plane to be at the equator of the vesicle,  the intensity field for the light arriving from membrane area elements is
\begin{align}
\label{eqn:intensity}
\mathcal{I}(r,\varphi)
	& \propto\iiint\mathrm{d}\Omega\;\mathcal{G}\left(r'\!\cos\theta'\right)\delta\left(r-r'\sin\theta'\right)\notag \\
	&\delta\left(r'\varphi'\sin\theta'-r\varphi\right)\,\delta\left(r' -  R\left[1+u(\theta'\!, \varphi')\right]\right),
\end{align} where $r$ and $\varphi$ are the polar coordinates in the equatorial plane, $\iiint\mathrm{d}\Omega \equiv \int^{\infty}_{0}\mathrm{d}r'\,{r'}^2 \!\int^{\pi}_{0} \mathrm{d}\theta'\sin\theta' \!\int^{2\pi}_{0}\mathrm{d}\varphi'\!$, and $\delta$ is the Dirac delta function. We assume that the membrane emits fluorescent light isotropically, so that the detected intensity is proportional to the projected membrane onto the focal plane, as sketched in Fig.~\ref{fig:1}(b). The intensity field $\mathcal{I}(r,\varphi)$ is  not directly experimentally observable, however its statistical moments are measurable quantities. Thus, the simplest way of extracting information is to analyze its first radial moment
\begin{equation}
  \label{eqn:first-mom}
  \rho_{\scriptscriptstyle\Delta}\!(\varphi) = \left.\int^\infty\limits_0 r\,\mathcal{I}(r,\varphi)\,\mathrm{d}r \middle/ \,\int\limits^\infty_0 \mathcal{I}(r,\varphi)\,\mathrm{d}r.\right.
\end{equation} This provides a closed contour, similar to $\rho_0(\varphi)$, in which the statistics of the contour represent a fluctuation spectrum analogous with (and asymptotically converging to) Eq.~(\ref{eqn:stduq}). As before, the deviations of $ \rho_{\scriptscriptstyle\Delta}(\varphi)$ are analyzed in Fourier space, where they are written as $\mu_q(t)$ \footnote{Change of notation from $u$ to $\mu$ reflects the fact that we are no longer transforming the membrane displacement at the equator, rather a moment of a intensity field.} and are non-dimensionalized by the mean radius $ R$. By defining $\bar{\mu}_q(t) \equiv \tau^{-1}\!\int^{\tau}_{0}\mathrm{d}t'\,\mu_q(t+t')$ as before to account for exposure time, the observed mean square amplitude can be exactly computed and has the same form as before
\begin{equation}
  \label{eqn:stdmu} \left\langle \bar{\mu}_q(t)\hspace{1pt}\bar{\mu}^{*}_q(t)\right\rangle_t = \sum_{n\,\geq\,q}\mathcal{L}^2_{n,q}\,\big\langle\!\hspace{1pt}\left|\hspace{1pt}\mathcal{U}^{q}_{n}\right|^2\!\big\rangle\,
 \frac{\tau^{2}_{n}}{\tau^{2}}\!\left(1-e^{-\tau/\tau_{n}}\right)^{\!2}\!\!,
\end{equation} where $\mathcal{L}_{n,q}$ is a function that depends on $\Delta$ (see Supplementary Materials \cite{Note1} for a complete derivation), namely
\begin{widetext}
\begin{equation}
 \mathcal{L}_{n,q} = \frac{\,1+(-1)^{n+q}\,}{\pi\hspace{1pt}I_0\left(\frac{1}{4\Delta^2}\right)e^{-\frac{1}{4\Delta^2}}}\int\limits^1_0\!\mathrm{d}\omega\;\mathcal{P}^{q}_{n}(\omega)\!\left[\frac{2\Delta^2 + \omega^2}{\Delta^2}-\frac{2\left(\Delta^2 + \omega^2\right)\text{erf}\left(\frac{1}{\Delta\sqrt{2}}\right)\,}{\,\sqrt{2\hspace{1pt}\pi\hspace{1pt}\Delta^2\left(1-\omega^2\right)\hspace{1pt}}\,I_0\left(\frac{1}{4 \Delta ^2}\right) e^{-\frac{1}{4 \Delta ^2}}}
 \right]\!e^{-\frac{\omega^2}{2\,\Delta^2}},
\end{equation}
\end{widetext} with $\mathcal{P}^{q}_{n}(\cos\theta) \equiv Y^{q}_{n}(\theta,\varphi=0)$, $\text{erf}$ being the error function, and $I_0$  the modified Bessel function of the first kind of order zero~\cite{Abramowitz1965}. In the limit of $\Delta\to0$, the fluctuation spectrum in Eq.~(\ref{eqn:stdmu}) is identically equivalent to~(\ref{eqn:stduq}).


\begin{figure}[b!]\includegraphics[width=\columnwidth]{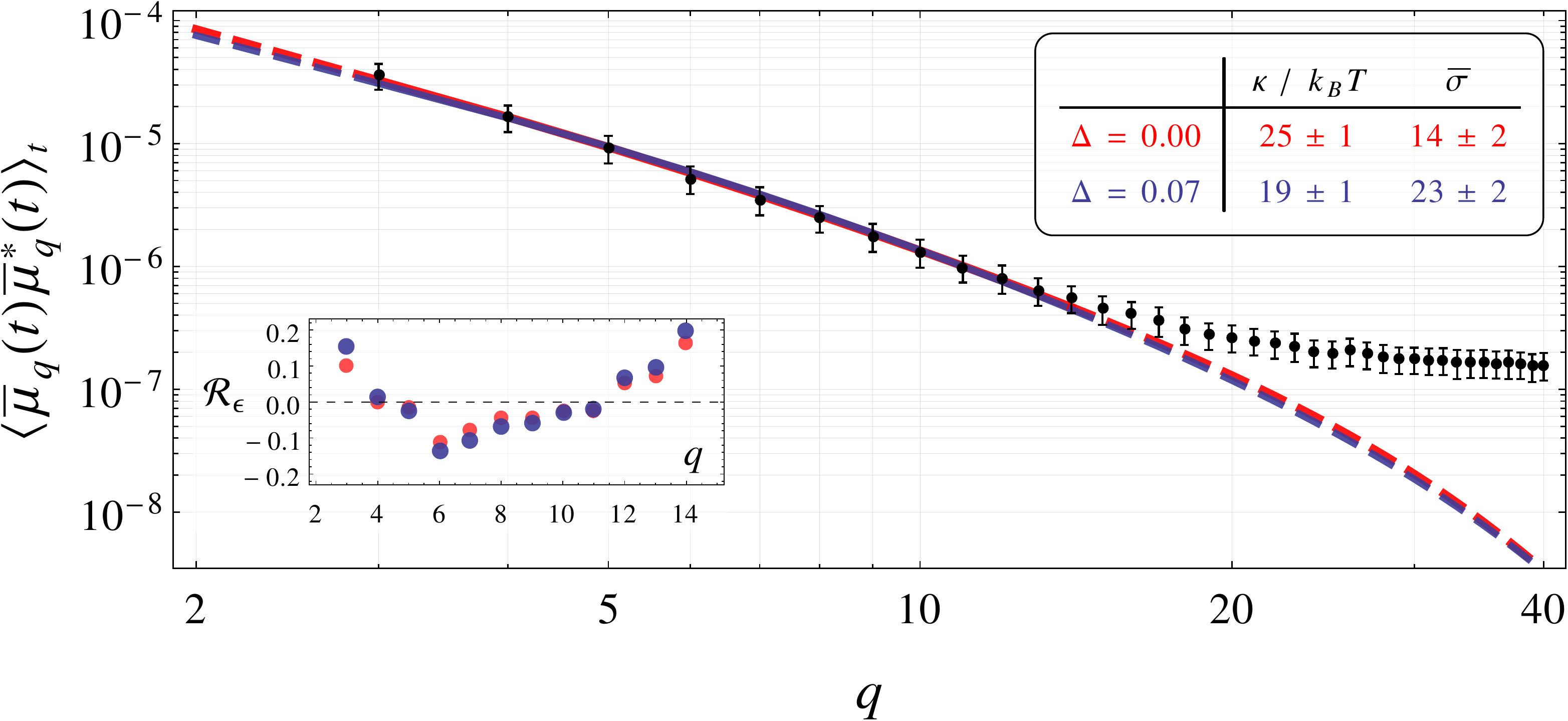}
\caption{\label{fig:2} (color on-line) Experimental fluctuation spectrum for a GUV ($ R \approx 10.2\;\mu\textnormal{m}$) imaged via confocal fluorescence microscopy, with known $\Delta=0.07$ and $\tau\approx1.2\;\textnormal{ms}$, plotted on a log-log scale. The best-fit lines for the models of Eq.~(\ref{eqn:stduq}) (i.e. assuming $\Delta=0$, incorrectly), and Eq.~(\ref{eqn:stdmu}) with $\Delta=0.07$ are shown by the red and blue solid curves, respectively. Although both fitting curves are of similar quality, importantly their best-fit values are significantly different, see the inset table (same colors). Dashed lines give the extrapolation of the curves outside the fitting range (see main text). The error-bars are standard errors in the mean, scaled up by a factor~10 to improve visibility. The inset plot shows the residuals $ R_\epsilon(q)$, normalized by their corresponding standard deviation.
}\end{figure}

\begin{figure}[b!]\includegraphics[width=\columnwidth]{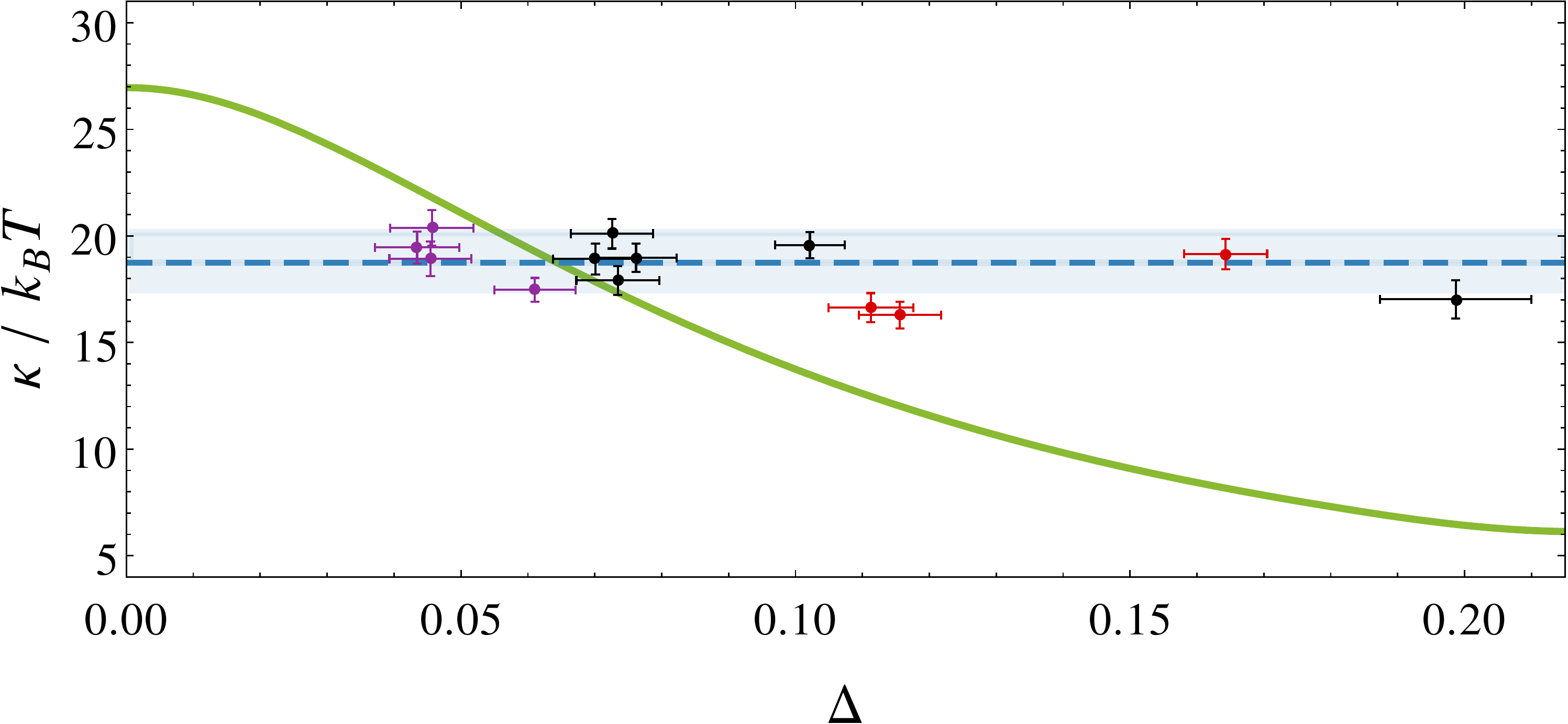}\caption{\label{fig:3} (color on-line) The inferred values of bending modulus~$\kappa$ from the fluctuation spectra of three GUVs of radii $6.4$~$\mu$m~(purple), $10.2$~$\mu$m~(black), and $16.7$~$\mu$m~(red), which are observed at different focal depths, $\Delta$. The~blue dashed line represents the estimate of $\kappa$ from the entire dataset $\mathcal{D}$, found by maximizing the posterior probability $\mathbb{P}\left(\kappa,\hspace{1pt}\bar{\sigma}_1, \bar{\sigma}_2, \bar{\sigma}_3\,\middle|\, \mathcal{D}\right)$, where $\bar{\sigma}_\ell$ is the corresponding reduced surface tension of $\ell$-th vesicle. This yields $\kappa=19\pm1\;k_B T$, where the 95\% confidence interval is shown as the blue shaded  band. The green curve shows the value of $\kappa$ one would find if the whole data~$\mathcal{D}$ is fitted at fixed $\Delta$. Large errors arise if $\Delta$ is not accounted for correctly. In particular, the value $\kappa\approx27\;k_B T$, found at $\Delta=0$, would be extracted from data if one neglected entirely the effect described in this work and used Eq.~(\ref{eqn:stduq}) to fit the experiments. The correction can be very important, and is essential to compare data from different $\Delta$, as can occur even in experiments with same optics but on GUVs of varying radius. }\end{figure}

The experiments were performed on GUVs prepared by means of electroformation as in~\cite{Parolini2015}, with DOPC and the~fluorescent lipid Texas Red 1,2-dihexadecanoyl-sn-glycero-3-phosphoethanolamine (DHPE) in proportions of 99.2\% and 0.8\%, respectively \cite{Note1}. To facilitate imaging, by reducing vesicle drift, the interior of GUVs is filled with a 197\,mM sucrose solution, whilst their exterior comprises of a 200\,mM glucose solution. Since the viscosities $\eta_\textnormal{in}$ and $\eta_\textnormal{out}$ are within 5\% of the viscosity of water at room temperature~\cite{Nikam2000}, they were both set to $\eta=1\,$mP\,s. The~microscopy was performed on a Leica TCS SP5 confocal scanning inverted microscope, using a HCX-PL-APO-CS $40.0\times$ oil immersion objective. Each vesicle is imaged in confocal epifluorescence mode~\cite{Mertz2009}. This allows control of the focal depth by varying the pin-hole size of the microscope. This imaging method is chosen instead of the more widely used phase contrast or bright field techniques, for precisely this reason. Here, the two-dimensional images are created through the raster scan, where the illumination spot is moved across the vesicle one  row at a time (of a pixel size $p\approx0.1\;\mu$m) at a line frequency $\nu=8$\,kHz. Therefore, this results in a very short acquisition time of a point of the projected membrane, $\tau\simeq\mathcal{W}/\hspace{1pt}v$, ranging between $1$--$2$\,ms in our experiments, where $v = p\hspace{1pt}\nu$ is the effective velocity of the line scanning front, perpendicular to the scan direction. However, this non-synchronous acquisition  leads to a cutoff in the mode spectrum, $q\!\lesssim\!\mathcal{Q}$, which is given by the condition that the mode half-lifetime associated to the highest (fastest relaxation) mode, from Eq.~(\ref{eqn:modelife}), should be longer than the time to scan across its wavelength: $\frac{1}{2}\,\tau_\mathcal{Q} = 2\pi R/(v\hspace{1pt}\mathcal{Q})$. Thus, we can assume that each portion of the raster scan of size $2\pi R/\mathcal{Q}$  samples membrane configurations from an equilibrium distribution, whereas the amplitudes on separate slices may have become temporally decorrelated (with our experimental settings and sample parameters we typically obtain  $\mathcal{Q}\approx20$). The fluctuation spectrum of the membrane is attained from such videos; the position of the contours in every frame is determined with sub-pixel precision by finding the position where each radial intensity profile has maximum correlation with a template (see Supplementary Materials \cite{Note1} for detailed data analysis). Then, we calculate the associated mean squared deviations of the contours in the $q$-space, $F_q(\Delta)$, averaged over $\sim1500$--$2000$ frames~\cite{Note1}. 

Using Eq.~(\ref{eqn:stdmu}), the best-fit values of $\kappa$ and $\bar{\sigma}$ to the experimental spectrum are found by means of a maximum posterior estimate \cite{Sivia2006}, assuming a uniform prior and that the measurement errors are independent and Gaussian \cite{Note1}; namely, we seek to minimize
\begin{equation}
 \label{eqn:fitting-chiSq-l}
 \chi^2_{\scriptscriptstyle\Delta}(\kappa,\bar{\sigma}) = \sum^{q_\textnormal{max}}_{\;q = q_\textnormal{min}}\left[\frac{F_q(\Delta) -  \big\langle\bar{\mu}_{q}(t)\,\bar{\mu}^{*}_{q}(t)\big\rangle_t}{\Sigma_q(\Delta)}\right]^{\!2}\!\!,
\end{equation} where $\Sigma_q(\Delta)$ is the standard error in the mean associated with $F_q(\Delta)$. Here, $q_\textnormal{min}$ and $q_\textnormal{max}$ define the lower and  upper bounds of the fitting range, respectively, with the former chosen to be $q_\textnormal{min} = 3$ \cite{Note1}. Due to the rapid convergence to zero of $\mathcal{L}^2_{n,q}$, the sum in Eq.~(\ref{eqn:stdmu}) is truncated at the mode $n = q+30$. On the other hand, the upper bound of the fitting range is selected as one that maximizes the posterior probability $\mathbb{P}\left(q_\textnormal{max}\,\middle|\,D_{\scriptscriptstyle\Delta}\hspace{-1pt}\right)$ based on data $D_{\scriptscriptstyle\Delta}\!=\{F_q(\Delta)\}_q$. This can be exactly computed by expanding $\chi^2_{\scriptscriptstyle\Delta}(\kappa,\bar{\sigma})$ to second order around the best-fit values of $\kappa$ and $\bar{\sigma}$ to the dataset $D_{\scriptscriptstyle\Delta}$ which yields
\begin{equation}
 \label{eqn:post-q-max}
 \mathbb{P}\left(q_\textnormal{max}\,\middle|\,D_{\scriptscriptstyle\Delta}\hspace{-1pt}\right) \propto \frac{e^{-\chi^2_\textnormal{min}/2}}{\sqrt{\det\!\big(\mathbf{H}_\textnormal{min}\big)}} \,\prod^{q_\textnormal{max}}_{q=q_\textnormal{min}}\!\frac{1}{\sqrt{\displaystyle2\hspace{1pt}\pi}\,\Sigma_q(\Delta)},
\end{equation} where $\chi^2_\textnormal{min} = \min\left[\chi^2_{\scriptscriptstyle\Delta}(\kappa,\bar{\sigma})\right]$, and  $\mathbf{H}_\textnormal{min}$ is the Hessian matrix  of Eq.~(\ref{eqn:fitting-chiSq-l}) evaluated at the best-fit values~\cite{Note1}. We also impose that $q_\textnormal{max}$ must be greater the crossover mode $q_c\equiv R \sqrt{\sigma/\kappa}$ \footnote{This crossover $q$-mode separates the regimes in which the membrane is mainly dominated by the surface tension term ($q\lesssim q_c$) and the bending rigidity term ($q\gtrsim q_c$). Thus, we require its value to lie within the fitting range.}, and less than the cutoffs $q_w\equiv R/\mathcal{W}$ and $\mathcal{Q}$, reflecting the optical and temporal resolution of our microscope, respectively, with $\mathcal{W}$ as the width of the lateral point spread function. In other words, the optimal fit is achieved when $\chi^2_{\scriptscriptstyle\Delta}(\kappa,\bar{\sigma})$ is minimal and simultaneously its upper bound $q_\textnormal{max}\in\left(q_c,\hspace{1pt}\min(q_w,\mathcal{Q})\right]$ maximizes the probability in Eq.~(\ref{eqn:post-q-max}). If $q_\textnormal{max}$ lies outside the interval defined above, then the dataset~$D_{\scriptscriptstyle\Delta}$ is rejected. See Supplementary Material for code and example files \cite{Note1}.

By imaging three GUVs with radii between $6$--$17\;\mu$m at various pin-hole sizes, the fluctuation spectrum associated with each $\Delta$ yields an individual estimate for the bending rigidity $\kappa$ and the surface tension. A systematic decrease in the inferred value of $\kappa$ is found when the data is fitted with the model in Eq.~(\ref{eqn:stdmu}) in comparison with the approach given by Eq.~(\ref{eqn:stduq}) that analyzes only equatorial fluctuations, as shown by the best-fit values in Fig.~\ref{fig:2}.  Using a maximum posterior estimate based on the data of all the vesicles (say~$\mathcal{D}$), including all their spectra at different values of $\Delta$, we conclude that $\kappa = 19\pm1$\;$k_B T$, as shown in Fig.~\ref{fig:3}. Here, since the surface tension is not a material parameter~\cite{Seifert1997}, the posterior probability is now a four-dimensional function $\mathbb{P}\left(\kappa,\hspace{1pt}\bar{\sigma}_1, \bar{\sigma}_2, \bar{\sigma}_3\,\middle|\, \mathcal{D}\right)$, with a different $\bar{\sigma}_\ell$ for each $\ell$-th vesicle. As when deriving Eq.~(\ref{eqn:fitting-chiSq-l}), we assume a uniform prior and that the measurement errors are independent and Gaussian; thus, maximizing this posterior probability function is equivalent to minimizing $\chi^2_\mathcal{D}\equiv\sum_{\Delta}\sum_{\ell}\chi^2_{\scriptscriptstyle\Delta}(\kappa,\bar{\sigma}_\ell)$, where each individual $q_\textnormal{max}$ is given by the maximum of Eq.~(\ref{eqn:post-q-max}). Note that if all the spectra are instead fitted using Eq.~(\ref{eqn:stdmu}), i.e.\ under the (incorrect) assumption $\Delta=0$, then one instead finds $\kappa = 27\pm1$\;$k_B T$, which is larger than the value found when using the correct, non-zero value(s) for~$\Delta$. To illustrate the dependence of the inferred values of $\kappa$ with the focal depth, the previous fitting procedure (that is, minimization of $\chi^2_\mathcal{D}$) is repeated at arbitrary non-zero values of $\Delta$ for all of the spectra in~$\mathcal{D}$, constructing an interpolation curve that is depicted by the green line in Fig.~\ref{fig:3}. This shows that the effect of the focal depth leads to a significant decrease in the value of $\kappa$ as $\Delta$ is increased. 

\begin{table}[t]
\centering
\begin{tabular}{lll}\hline\hline\\[-10pt]
Experimental method			& &	\hspace{0.78cm}$\kappa\hspace{1pt}/\hspace{1pt}k_BT$\;\\[2pt]
\hline\hline\\[-8pt]
Flicker spectroscopy of GUVs						& & \\
  \qquad Literature values \cite{Brown2011,Nagle2013,Gracia2010}:	& & \hspace{0.8cm}$27\pm2$\;\\[1pt]
  \qquad Present work (optical projection):				& & \hspace{0.8cm}$19\pm1$\;\\[1pt]
X-ray scattering on bilayer stacks \cite{Kucerka2006, Kucerka2005,
Pan2008a,Pan2008}: 							& & \hspace{0.8cm}$18\pm2$\;\\[1pt]
Pulling membrane tethers \cite{Tian2009, Sorre2009}:			& & \hspace{0.8cm}$19\pm2$\;\\[1pt]
Micropipette aspiration of GUVs \cite{Rawicz2000}:			& & \hspace{0.8cm}$19\pm2$\;\\[2pt]
\\[-10pt]\hline\hline
\end{tabular}
\caption{\label{table:kappa-by-other-exp} Measurements of $\kappa$ for  DOPC lipid membranes from different experimental techniques, including the estimate of the present work and the previous literature values (shown as the mean estimates from each cited work).}
\end{table}

In summary, we propose a model for flickering spectroscopy based on a projection of shape fluctuations that accounts for the focal depth of the microscope. Our approach brings the mechanical parameters estimated from flickering experiments into full agreement with those found by other experimental methods \cite{Nagle2013}, such as X-ray scattering on membrane stacks~\cite{Fragneto2003, Daillant2005, Hemmerle2012}, micropipette aspiration techniques~\cite{Evans1990, Zhelev1994, Rawicz2000, Fournier2001, Henriksen2004}, and methods of pulling membrane nanotubes from GUVs~\cite{Borghi2003, Cans2003, Heinrich1996, Koster2005, Cuvelier2005, Hochmuth1982a, Hochmuth1982b}, see Table~\ref{table:kappa-by-other-exp}. Previously, the literature presented a puzzle \cite{Nagle2013}: the inferred $\kappa$ from the shape analysis of GUVs were larger than those obtained from other techniques. Flickering spectroscopy has several advantages over the aforementioned techniques; it relies on general purpose and easily accessible equipment, it is non-invasive, and can be integrated into microfluidic devices, avoiding the manipulation of GUVs. These advantages make it as a popular methodology for the mechanical analysis of membranes. The theoretical framework and data analysis procedure presented here raises its degree of accuracy to that of more involved methods, ultimately enabling low cost and reliable micromechanical characterisation of membranes.

\begin{acknowledgments}
We acknowledge  stimulating discussions with P.\,Sens, M.\,Polin, and A.\,T.\,Brown, and the sample preparation support from L.\,Parolini. This research work is funded by EPSRC under grants EP/I005439/1~(M.S.T.) and EP/J017566/1~(P.C.), and Project SPINNER~2013, Regione Emilia-Romagna, European Social Fund~(D.O.).
\end{acknowledgments}


%

\end{document}